\documentclass{elsart}  

\usepackage{lmodern}
\usepackage{epsfig,amsmath,amsfonts,amssymb,setspace,multirow,textcomp}
\usepackage[T1]{fontenc}
\usepackage{times}
\usepackage{fancyhdr}
\usepackage{type1cm}
\usepackage{wrapfig}
\usepackage{graphics}
\usepackage{listings} 
\usepackage{color}
\usepackage{ragged2e}
\usepackage{txfonts}
\usepackage{subfig}
\usepackage[countmax]{subfloat}
\usepackage[authoryear]{natbib}

\begin{document} 

\begin{frontmatter}
   \title{Physical properties of solar polar jets: \\ a statistical study with Hinode XRT data}

\author{A.~R.~Paraschiv}
\address{  INAF-Turin Astrophysical Observatory, via Osservatorio 20, 10025 Pino Torinese, Italy\\
   Institute of Geodynamics $"$Sabba S. Stefanescu$”$ of the Romanian Academy,
19-21, str. J.L. Calderon, Bucharest, Romania}
\ead{paraschiv.alin.razvan@gmail.com;}

\author{ A.~Bemporad}
\address{  INAF-Turin Astrophysical Observatory, via Osservatorio 20, 10025 Pino Torinese, Italy}   
\ead{bemporad@oato.inaf.it}   
   
\author{A.~C.~Sterling}         
\address{NASA/Marshall Space Flight Center,Huntsville, Al 35812, USA
\ead{Alphonse.Sterling@nasa.gov}}

   \date{\today}

  \begin{abstract}

   The target of this work is to investigate the physical nature of polar jets in the solar corona and their possible contribution to coronal heating and solar wind flow based on the analysis of X-ray images acquired by the Hinode XRT telescope. We estimate the different forms of energy associated with many of these small-scale eruptions, in particular the kinetic energy and enthalpy. Two Hinode XRT campaign datasets focusing on the two polar coronal holes were selected to analyze the physical properties of coronal jets; the analyzed data were acquired using a series of three XRT filters. Typical kinematical properties (e.g., length, thickness, lifetime, ejection rate, and velocity) of 18 jets are evaluated from the observed sequences, thus providing information on their possible contribution to the fast solar wind flux escaping from coronal holes. Electron temperatures and densities of polar-jet plasmas are also estimated using
ratios of the intensities observed in different filters. We find that the largest amount of energy eventually provided to the corona is thermal. The energy due to waves may also be significant, but its value is comparatively uncertain. The kinetic energy is lower than thermal energy, while other forms of energy are comparatively low. Lesser and fainter events seem to be hotter, thus the total contribution by polar jets to the coronal heating could have been underestimated so far. The kinetic energy flux is usually around three times smaller than the enthalpy counterpart, implying that this energy is converted into plasma heating more than in plasma acceleration. This result suggests that the majority of polar jets are most likely not escaping from the Sun and that only cooler ejections could possibly have enough kinetic energy to contribute to the total solar wind flow.

  \end{abstract}

   \end{frontmatter}

%

\section{Introduction}

\indent\indent  Coronal jets are small collimated ejections of plasma observed in ultraviolet/x-ray imagers and in white-light coronagraph images \citep[see][]{stcyr1997}. The kinematics of polar jets were first studied by \citet{wang1998} during solar minimum activity, using images from the Large Angle and Spectrometric Coronagraph \citep[LASCO;][]{brueckner1995} and the Extreme ultraviolet Imaging Telescope \citep[EIT;][]{delaboudiniere1995} onboard the SOlar and Heliospheric Observatory \citep[SOHO;][]{domingo1995}. The authors analyzed 27 correlated events and derived outflow speeds of $\sim 250 km\cdot s^{-1}$. More recently, \citet{cirtain2007} used Hinode \citep[][]{kosugi2007} observations of polar coronal holes to reveal that x-ray jets have two distinct velocities: one near the Alfv\'en speed ($600-800\text{ }km\cdot s^{-1} $) and another near the sound speed ($\sim 200 km\cdot s^{-1}$). This indicated that the jets may contribute to the high-speed solar wind. These observations also demonstrated that jets are transient phenomena that occur at much higher rates than large-scale events, such as flares and coronal mass ejections.

On the other hand, the morphological properties of coronal jets have been observed in soft X-rays by the Yohkoh mission \citep[][]{ogawara1991}. \citet{shimojo1996} and \citet{shimojo2000} analyzed jets using data from the Soft X-Ray Telescope \citep[SXT;][]{tsuneta1991} onboard Yohkoh. They saw X-ray jets mainly near or in active regions and suggested that they are produced by reconnection occurring at the footpoint of the jets. \citet{culhane2007} found that, in time, jet plasma cools and falls back to its original site, as is consistent with the existing models that involve magnetic reconnection. The study of \citet{savcheva2007} based on observations from the X-Ray telescope \citep[XRT;][]{golub2007} onboard Hinode showed that around 60 x-ray jets per day occur on average inside the polar coronal holes. The apparent outward velocity, the height, the width, and the lifetime of the jets were measured, and a statistical study of jet transverse motions and backflows was also presented. Recently, \citet{chandrashekhar2014} and \citet{Chandrashekhar20142} studied the dynamics of two jets observed with the Hinode/XRT and Hinode/EIS and confirmed the drift motions that appear in the upward movement of the event, suggesting that the jet material is undergoing cooling and falling back. 

Correlations between jets and other signs of solar activity have been studied, for instance, by \citet{raouafi2008}, who focused on the temporal evolution of and relationships between polar coronal jets and polar plumes, and showed that $\sim 90\%$ of the observed jet events are associated with polar plumes. \citet{filippov2009} discuss the formation of jets and proposed scenarios that explain the main features of the events: the relationship with the expected surface magnetism, the rapid and sudden radial motion, and possibly the heating, based on the assumption that the jet occurs above a null point of the coronal magnetic field. Using the Solar Terrestrial Relations Observatory \citep[STEREO;][]{kaiser2008} twin inner coronagraphs \citep[COR1;][]{thompson2003} \citet{nistico2009} studied various aspects of jets, including their correlation with underlying small scale chromospheric bright points. More recently, \citet{pucci2012} have presented an analysis of the correlation of X-ray bright points and jets observed by Hinode/XRT and concluded that most of the jets occurred in close temporal association with the brightness maximum in bright points. A dichotomy of polar X-ray jets was proposed by \citet{moore2010,moore2013}. They observed that about half of their sampled events fit the standard reconnection picture for coronal jets, and the rest correspond to another type. The non-standard jets (which they call "blowout jets") are described as having a jet-base magnetic arch that undergoes blowout eruptions, similar to, but smaller than those that produce major coronal mass ejections. The authors also propose possible correlations between jets and Type-II spicules and macrospicules and conclude that the combined energy output could contribute significantly to the heating of the corona.

The physical parameters of jet plasma have been derived, for instance, by \citet{shimojo2000} with SXT data on Yohkoh. More recently, \citet{nistico2011} have described the typical physical characteristics of coronal jets observed by the SECCHI instruments of STEREO spacecraft obtaining a temperature determination for the jet plasma. Their results show that jets are characterized by electron temperatures ranging between 0.8 MK and 1.3 MK, similar to results by \citet{young2014}, who studied the link between a blowout jet and its base brightpoint. \citet{pucci2013} analyzed the difference between a standard and a blowout jet seen from Hinode-XRT. They aimed at inferring differences in physical parameters corresponding to the two categories proposed by \citet{moore2010}. \citet{pucci2013} conclude that their blowout jet was hotter, had a higher outflow speed, and was rooted in a stronger magnetic field region compared to their standard jet.

In our previous work \citep{paraschiv2010}, we identified and studied white-light jets observed by the SECCHI-COR1 white light coronagraph. We identified more than 10~000 white-light jets spread across the entire solar limb at the minimum of solar activity between 2007 and 2008. The identified jets originated in regions inside coronal holes but also in regions of the quiet Sun. A subsamble of the events were considered for correlation with bright points on the solar disk. The association was done based on the changes (morphological, brightening, disappearing, etc.) of the bright points as observed in Extreme Ultraviolet Imager \citep[EUVI;][]{wulser2004} $195\mathring{A}$ images at the time when the jet was observed in COR1. In \citet{paraschiv2010}, we suggested that the number of jets with high outflow speeds is large enough to contribute significantly to the solar wind, and we concluded that these results may point to the mechanism responsible for the expansion of the jet, namely a pressure-driven expansion of the plasma. We later expanded the study \citep{paraschiv2012} in order to discuss the possible contribution of coronal jets to the solar wind flux, by estimating the ejected particle mass flux of a selected subsample of the above-mentioned coronal jets.
\begin{wrapfigure}{r}{0.55\textwidth}
\centering
\includegraphics[width=\linewidth]{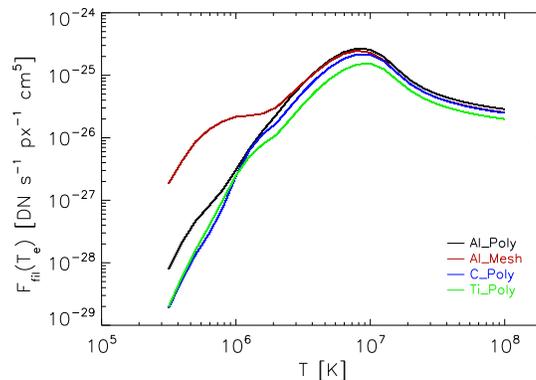}
\caption{Total XRT temperature response, given for the X-ray focal-plane filters we used. Each curve plots the combination of the total instrument response as a function of wavelength, assuming a coronal plasma emission model (ATOMDB/APEC). The full filter response curves are available in \citet[][]{narukage2014}.}
\label{fig:tempresp}
\end{wrapfigure}
In the present work we want to extend our previous analyses to X-ray data and to derive plasma physical parameters on a statistically significant sample of jets. The data that were used, along with the physical phenomena and models and assumptions involved in the analysis, are described in Section 2. The third section depicts the methods used, computations, and results obtained: the jet's geometrical and kinematic properties, jet temperature, and density determinations are described. An estimate of the energy budget for coronal jet eruptions is also presented. In the fourth Section, discussions, results and conclusions are presented. Comparisons to previous studies are given and conclusions about possible contribution to coronal heating are made.

\section{Data selection \& calibration}

\indent\indent All the data analyzed in this work were acquired by the XRT instrument onboard the Hinode spacecraft. The XRT is a modified Wolter I telescope that uses grazing incidence optics to image the solar corona's emissions (T$>\sim$1MK) with an angular resolution of 1 arcsec per pixel at the CCD. The telescope has a maximum imaging field of view of 34 arcminutes. It uses several filters tuned
to differing wavelengths, allowing sampling of different regions of the corona. \citet{golub2007} gives full details for the instrument.

Figure \ref{fig:tempresp} shows the response functions, $F_{fil}(T_{e})$, for the filters used here as functions of temperature. XRT operates on a mission schedule and changes its pointing accordingly. For the purpose of this analysis, we had to search for suitable sets of data. Instrument pointing in polar coronal holes was required for a good (via a better signal-to-noise ratio) differentiation between the jets and the coronal background. Full resolution of 512 $\times$ 512 pixels with a spatial resolution of 1.02 arcsec/px and adequate image cadence (temporal resolution) of no longer than $\sim 2$ min were used as mandatory data conditions for this study. The data also needed to be available in multiple filter combinations in order to apply the filter ratio technique used to determine parameters of the jet plasma, such as the temperature, density, and energy budget of events. 

\begin{wrapfigure}{r}{0.55\textwidth}
\centering
\includegraphics[width=\linewidth]{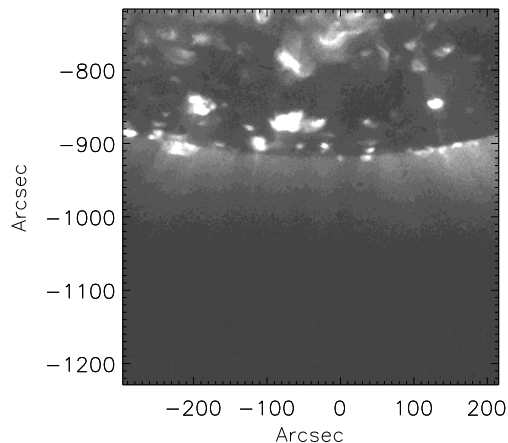}
\caption{Example of XRT data field of view for the observations acquired on 26 Jan 2009.}
\label{fig:xrtfov}
\end{wrapfigure}

We selected the following datasets:
\begin{itemize}
\item 1. Dataset acquired on 6 July 2008, containing approximately four hours (10:36 to 14:40) of continuous observations of the north polar coronal hole. Data were available at full resolution, in three filters (Al\_Poly, Al\_Mesh, C\_Poly) with $\sim1.5$ minute cadence between each three-filter exposure set. 
\item 2. Dataset acquired on 26 January 2009, containing approximately five hours (07:02 to 11:54) of continuous observations of the south polar coronal hole. Data were available at full resolution, in three filters (Al\_Poly, Al\_Mesh, Ti\_Poly) with approximately a two-minute cadence between each three-filter exposure set. 
\end{itemize}

An example of a typical XRT instrument field of view is shown in Figure \ref{fig:xrtfov}. Data were retrieved online through the archive search page of the Hinode Science Data Center (SDC) Europe (http://sdc.uio.no/search/). All the data were analyzed using the Solarsoft (SSW) package, written in IDL language. The xrt\_prep.pro procedure that was used to calibrate raw XRT data from Levels 0 to 1 mainly performs the following actions: 1) correction for undersaturated and oversaturated pixels, 2) removal of spikes that occur due to the radiation belt and/or cosmic rays, 3) correction of images for contaminating material that has accumulated on the instrument, 4) co-alignement of the images by applying an orbital drift correction to reduce the effects of satellite jitter, 5) normalization of the output image in DN s$^{-1}$, and 6) subtraction of instrumental dark noise using dark frames acquired close in time to each of the two datasets. Correction for contamination is very important; since Hinode's launch, contaminating material has accumulated on the XRT CCD and focal plane filters, causing a decrease in sensitivity. The effect of contamination is wavelength-dependent, and the thickness of the contaminant deposited on the focal plane filters is different, so that contamination affects the response of each XRT filter differently. See \citet{narukage2011} for further details on analysis and modeling of contamination effects on XRT data and responses.

\section{Data analysis}

\subsection{Geometrical \& kinematical parameters}

\indent\indent A total of 18 jet events were identified and selected by visual inspection of the approximately nine hours of observations that we analyzed. Image sequences were employed to derive many different geometrical and kinematical properties of these events. For each event, we first selected the frame with the jet at its maximum elongation prior to when it started to dissipate. A rotation was then applied so that the jet was vertical in our analysis window, and the region containing that vertical jet was preselected in a rectangular box. Then each event was isolated based on its intensity profile along the direction perpendicular to the jet's motion; the resulting jet widths ranged from 3 to 12 pixels.

\begin{figure}[!t]
\centering
\includegraphics[width=\linewidth]{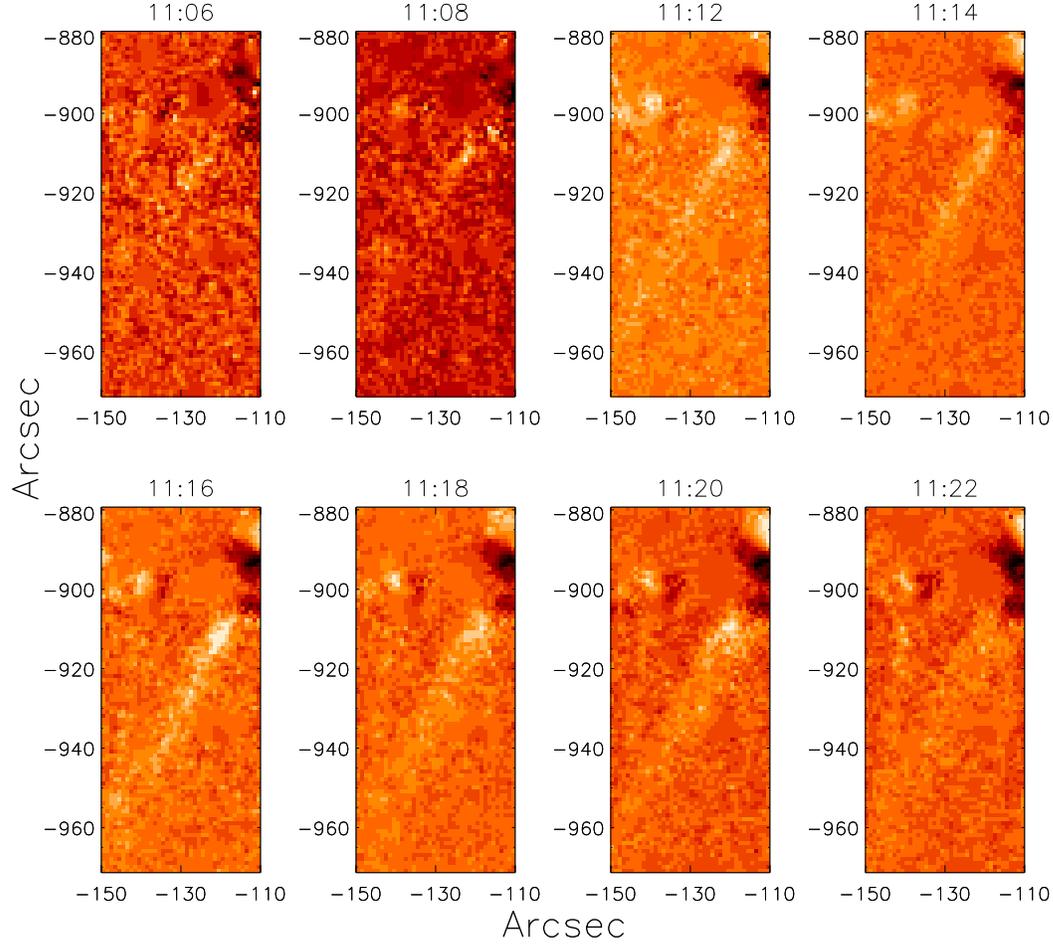}
\caption{Example of jet eruption in a sequence of base difference images (prior to rotating the frame for analysis; see text) showing the rising phase, the main phase, and the decay phase of a jet event. The outflow speed can be computed by time tracking the leading edge of the eruption. This sequence was acquired between 11:06 UT and 11:26 UT on 26 Jan 2009 with the Al$\_$Poly filter.}
\label{fig:jetseq}
\end{figure}

Jet outflow speeds $\mathtt{v}_{out}$ were estimated by summing each rectangular area over the direction perpendicular to the jet (jet width), providing the jet intensity distribution along its axis, and then by tracking for each event the position of the leading edge, identified as the location where the intensity dropped below a fixed threshold, in successive frames. An example of a sequence acquired during one of the jets analyzed here is shown in Figure \ref{fig:jetseq}. Because all the jets we selected are observed off-limb, the plasma mainly propagates over the plane of the sky, so that the possible projection effects due to jet inclinations are minimized in the velocity measurement. For each individual event, data are available as independent series for each of the three filters, thus three outflow speed values were obtained. The speeds averaged over all the events are $\mathtt{v}_{Al\_Poly}=157\pm 31$ km s$^{-1}$,$\mathtt{v}_{Al\_Mesh}=163\pm 25$ km s$^{-1}$, and $\mathtt{v}_{C\_Poly \, and\,  Ti\_Poly}=156\pm 25$ km s$^{-1}$; different outflow speeds for all 18 events are provided in the jet parameter table (see Appendix). Computed errors are standard $1\sigma$ deviations. More significant errors are due to uncertainties in the exact identification of the location of the jets' leading edges, and the estimated resulting uncertainties in the outflow speed values are in the range of about 20\%-25\%. By summing each rectangular area over the direction parallel to the jet axis, we get a profile of the average jet intensity across its width: a $2\sigma$ dispersion from the location of the brightness maximum was selected as being the value of the typical jet width $d$. Values of the widths of the jets are also provided in the jet parameter table (see Appendix); the average widths of all of the jets is $d\simeq 8\pm1$ pixel $(5900\pm750$ km).

\subsection{Plasma temperatures}

\indent\indent Using XRT data acquired with multifilter sequences, we can measure the electron temperature of the jet plasma via the filter-ratio technique. As mentioned, filters on the XRT telescope are characterized by a response function $F_{fil}$ that depends on the temperature usually given in the units of $[DN\cdot s^{-1}\cdot pixel^{-1}]$ for a unit column emission measure (CEM) measured in $[cm^{-5} ]$. The measured intensity for a particular filter $I_{fil}$ can be written as an integration along the line of sight (LOS) as
\begin{figure*}[t]
\centering
\includegraphics[width=0.8\linewidth]{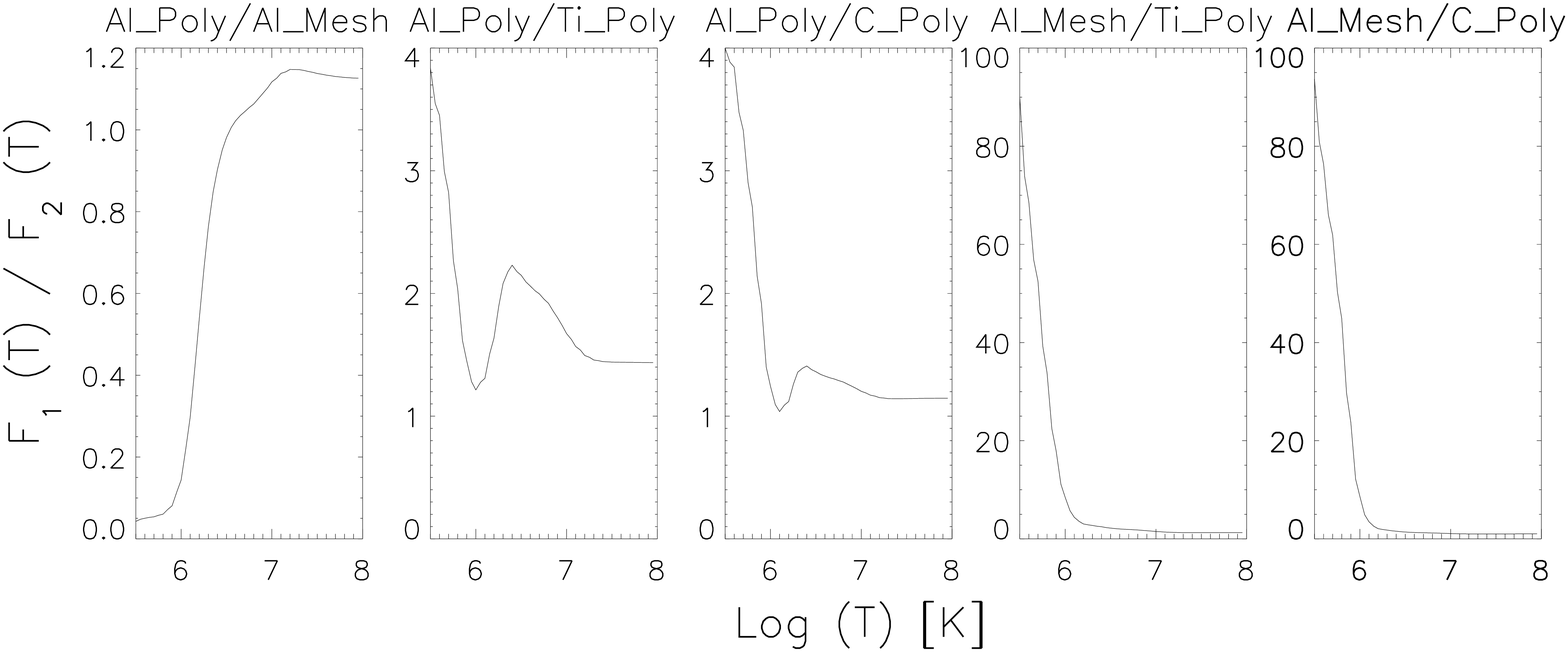}
\caption{Ratios computed with the filter response functions. }
\label{fig:filtratio}
\end{figure*}

\begin{equation}
I_{fil} = \int_{LOS} F_{fil}(T_{e}) n^{2}_{e} dl  \quad [DN \cdot s^{-1} \cdot pixel^{-1} ].
\label{eq:ifilter1}
\end{equation}

The above equation shows that the measured intensities depend on the filter response, which is in turn a function of the plasma electron temperature $(T_{e})$. The filter response function also depends upon the abundance of the elements whose lines contribute to the emission measured by the instrument.

The emitting plasma is certainly not isothermal along the LOS. Nevertheless, because the electron temperature distribution along the line of sight is unknown, some assumptions are needed to apply the filter ratio technique. In particular, if the emission due to the background corona is subtracted so that the emission due to the jet plasma alone is isolated, and if we assume the jet is isothermal along the LOS \citep[see][]{moreno2013}, the above equation can be rewritten as

\begin{equation}
I_{fil} = F_{fil}(T_{e}) \int_{l} n^{2}_{e} dl  \quad [DN \cdot s^{-1} \cdot pixel^{-1} ].
\label{eq:ifilter2}
\end{equation}

 The measured intensity also depends on the plasma column emission measure, $CEM=n^{2}_{e} \cdot l$ [cm$^{-5}$], where $n_{e}$ [cm$^{-3}$] is the electron number density. For this, we considered the jets to have a cylindrical geometry and thus the depth $l$ is assumed to be equal to the projected width $d$ on the plane of the sky of the studied events.
 
Under our isothermal assumption the column emission measure is the same for each filter, and so the ratio $R$ between the intensities measured in the same jet with two different filters (after subtraction of coronal background) solely depends on the electron temperature:
\begin{equation}
R(T_e) = \frac{I_1(T_{e})}{I_2(T_{e})} = \frac{F_1(T_{e})}{F_2(T_{e})}=\frac{DN_1/t_1}{DN_2/t_2} \Rightarrow T_{e}=R^{-1}\left(\frac{DN_1/t_1}{DN_2/t_2} \right) 
  \label{eq:filtratio}
  \end{equation}

Equation \ref{eq:filtratio} describes the direct dependence between the emitting plasma electron temperature and the DNs detected for one jet with two different filters and different exposure times $t_i$. Given the observed ratio and the theoretical one, it is thus possible to derive the plasma temperature for each pixel: different ratios employed for this analysis are shown in Figure \ref{fig:filtratio} as functions of temperature. As can be seen in this figure, for some very limited ranges of filter ratios the temperature is not uniquely determined, because two different temperatures are associated with the same single value of filter ratio. We handled these few double-valued situations by selecting the temperatures closer to the average value obtained from the other filters, which is in general the lowest possible temperature.

\begin{figure*}[t]
\centering
\captionsetup[subfigure]{labelformat=empty}

\subfloat[a]{
\label{fig:jettemp}
\includegraphics[width=0.45\linewidth]{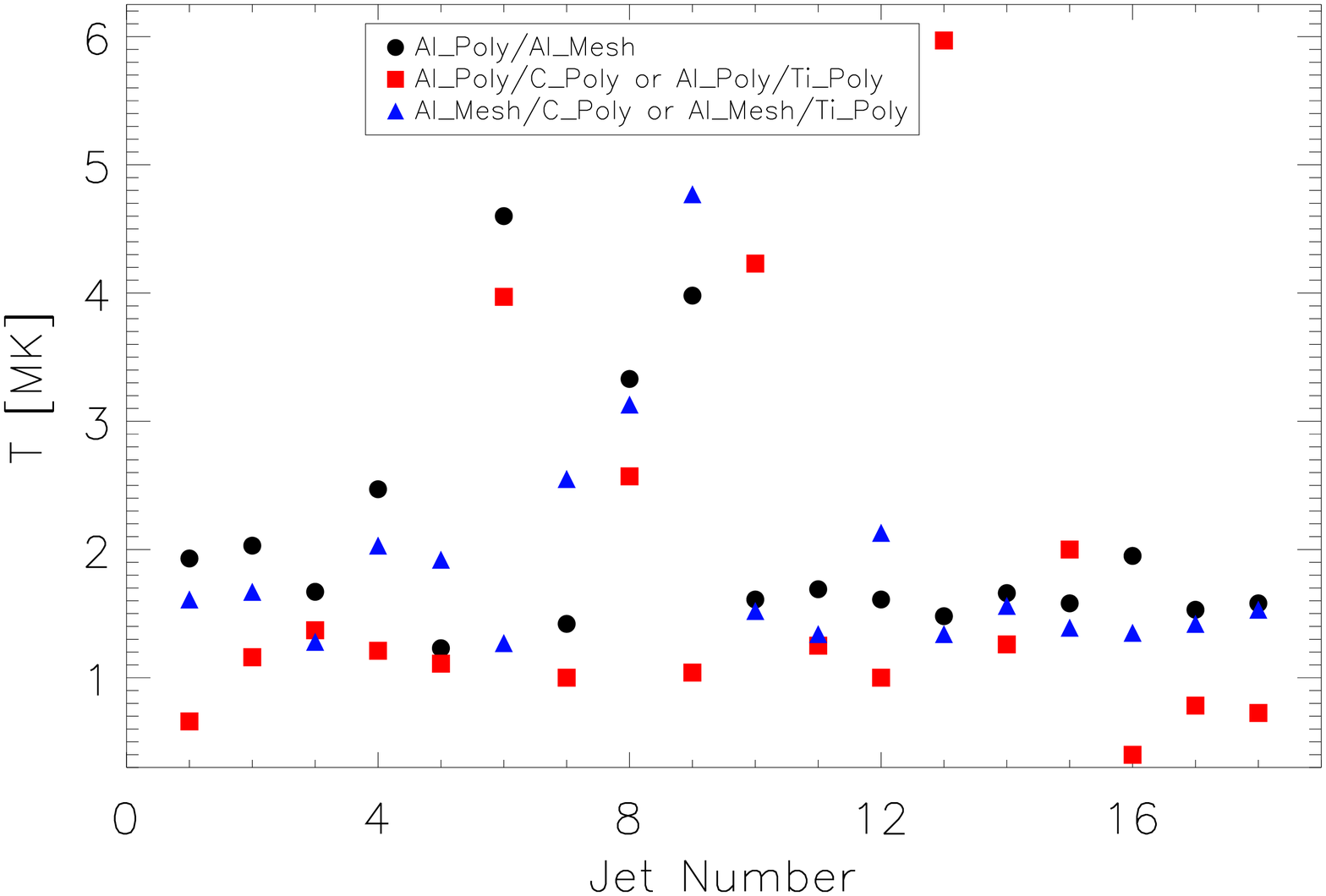}}
\hspace{0.05\linewidth}
\subfloat[b]{
\label{fig:jetdens}
\includegraphics[width=0.45\linewidth]{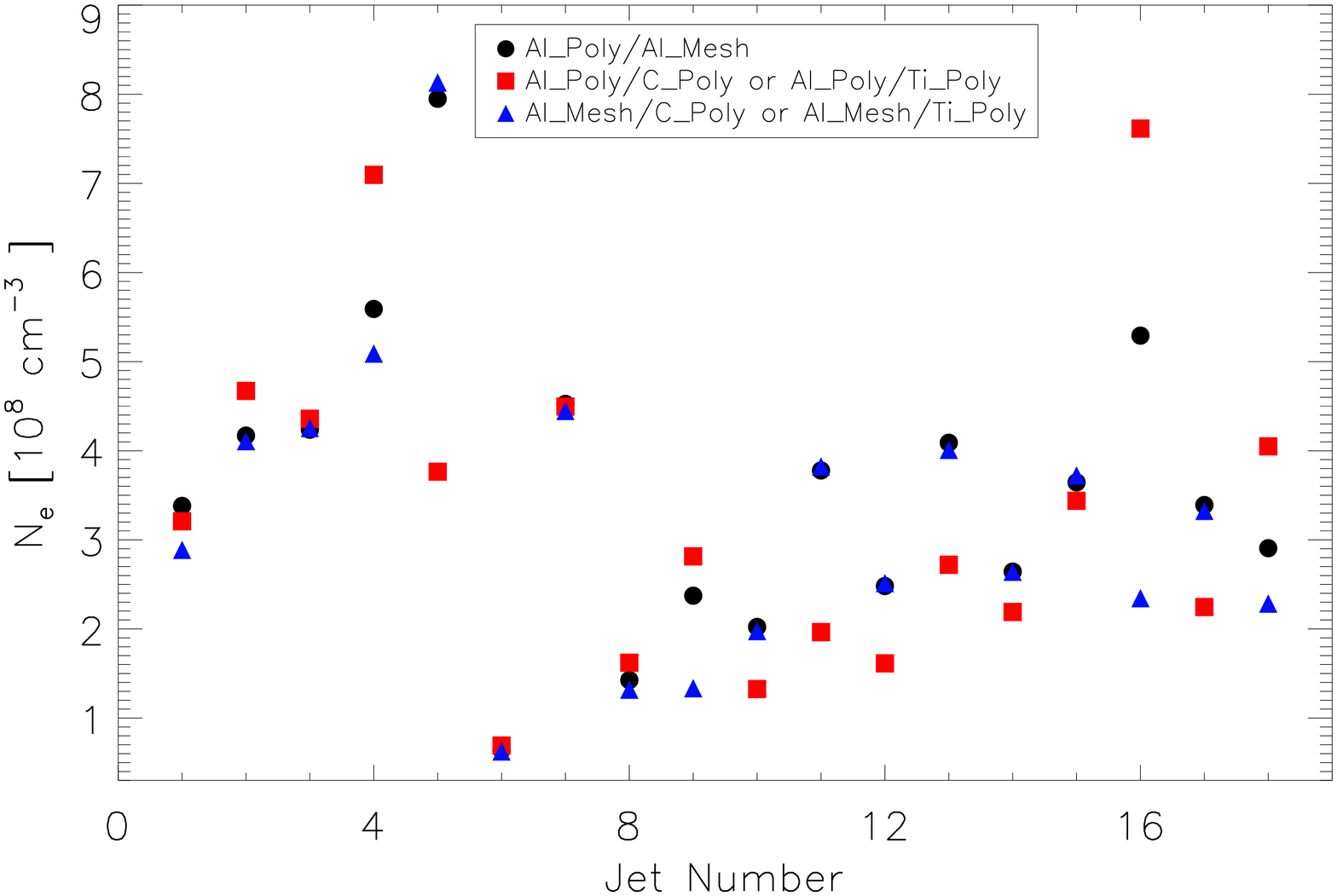}}

\caption{Jet parameters derived using the three filter ratio combinations. Left panel (a): Jet temperatures ($y-$axis) derived for the 18 jets studied here ($x-$axis) using different filter ratios. Right panel (b): Jet densities ($y-$axis) derived, using Eq. \ref{eq:no4_7}, for all events ($x-$axis) using different filter ratios. Different filter pairs are depicted using different symbols and colors (as shown in the legend). Computed values usually correspond closely between the Al\_Poly/Al\_Mesh, Al\_Mesh/C\_Poly, and Al\_Mesh/Ti\_Poly ratios, while values derived from the Al\_Poly/C\_Poly or Al\_Poly/Ti\_Poly ratio are somewhat different.} 
\end{figure*}

In summary, the following steps were followed in order to derive the average jet temperature. An intensity profile along the jet was created at the moment of its maximum extent, together with a corresponding intensity profile for the solar corona obtained at the same spatial location and the same range of altitudes with frames acquired just before the jet's start time. The latter profile was fitted with a second-order polynomial and removed from the intensity profile along the jet, thereby removing the estimated background and foreground coronal emissions from it without increasing the noise. From the ratio between the intensities observed with different filters, the temperature was then calculated, using the curves of Figure \ref{fig:filtratio}, for each ten-pixel interval along the jet. The resulting temperatures usually had roughly constant values along the length of the jet; as a result, by doing a global average over the entire jet, we get the average temperature of each jet (Figure \ref{fig:jettemp}).

 Only three filter ratios were computed for each event owing to the different available filter combinations of the two used datasets. Some ratios give more reliable results than others: The derived temperatures are in general quite similar for the Al\_Poly/Al\_Mesh, Al\_Mesh/Ti\_Poly, and the Al\_Mesh/C\_Poly ratios, while the Al\_Poly/C\_Poly and Al\_Poly/Ti\_Poly ratio gives slightly different results. This is partly because some filters are more affected by contamination than others, and also because the ratio of the response functions of the Al\_Poly/C\_poly and Al\_Poly/Ti\_Poly filters is too flat in that range of temperatures (see Figure \ref{fig:filtratio}). Considering that there were substantially larger variations in temperatures derived with the Al\_Mesh/C\_Poly or Al\_Mesh/Ti\_Poly ratios compared to the Al\_Poly/Al\_Mesh ratio, we concluded that the latter gave slightly better results. This ratio is then the only one we used in further measurements. The estimated uncertainties computed using the error estimation given by \citet{narukage2011} are on the range 10\% - 15\%. The average jet temperatures are $1.8\pm0.2 \cdot 10^6$ K, temperatures measured for single events are shown in Figure \ref{fig:jettemp}: these results will be discussed in the next section.

\subsection{Plasma densities}

\indent \indent Plasma temperatures obtained with the filter ratio technique were used in turn to derive an estimate of the plasma density. Values for the densities of both jet and surrounding coronal plasmas are computed at the moment of maximum brightness during the jet evolution. In this work we define the jet density $n_{e\_jet}$ as the quantity to be added to the density of the surrounding corona $n_{e\_cor}$ in order to get the density measured from the observed X-ray emission, $n_{e\_obs}=n_{e\_cor}+n_{e\_jet}$. The observed jet intensity in a given filter (under the hypothesis of nearly constant temperature along the LOS) after subtraction of the coronal background is given by Eq. \ref{eq:ifilter2} \citep[see also][]{pucci2013}.
As previously mentioned, the assumption that the jet has a simple cylindrical form was adopted, so $l$ is assumed to be equal to the observed width $d$ of the jet. If along the LOS no significant changes of the quantity $F_{fil}(T_{e})$ occur with respect to the quantity $n_e^2$, we can separate the above equation into two components: the emission $I_{jet}$ observed at the location of the jet, and the emission $I_{cor}$ observed at the same location in the corona but before the jet. The $I_{cor}$ quantity will be given simply by
\begin{equation}
I_{cor} = F_{fil}(T_{e})\int_{-\infty}^{\infty} n^{2}_{e\_cor} dz,
\label{eq:no4_5}
\end{equation}
while the $I_{jet}$ quantity can be written as
\begin{equation}
\resizebox{0.93\hsize}{!}{$I_{jet} = F_{fil}(T_{e})\left[ \int_{-\infty}^{-l/2} n^{2}_{e\_cor} dz + \int_{-l/2}^{l/2} (n_{e\_cor}+n_{e\_jet})^2 dz+ \int_{l/2}^{\infty} n^{2}_{e\_cor} dz \right].$}
\label{eq:no4_6}
\end{equation} 

Because in our analysis the background pre-jet intensity was subtracted, the resulting intensities correspond to the quantity $I_{jet}-I_{cor}$. Thus for the two filters that were used we can write
\begin{equation}
\begin{array}{llcc}
(I_{jet}-I_{cor})_{Al\_Poly} &= F_{Al\_Poly}(T_{e})\cdot l \cdot (n_{e\_jet}^2+2\cdot n_{e\_jet} \cdot n_{e\_cor})\\
(I_{jet}-I_{cor})_{Al\_Mesh} &= F_{Al\_Mesh}(T_{e})\cdot l \cdot (n_{e\_jet}^2+2\cdot n_{e\_jet} \cdot n_{e\_cor}).
  \label{eq:no4_7}
\end{array} 
\end{equation}
Given the observed intensities, the only two unknown quantities in the above two equations are the densities $n_{e\_jet}$ and $n_{e\_cor}$, which can then be determined. Densities derived for all 18 events are shown in Figure \ref{fig:jetdens}; average values of $n_{e\_jet}=1.5\pm0.1 \cdot 10^8$ cm$^{-3} $ and $n_{e\_cor}=1.6\pm0.2 \cdot 10^8$ cm$^{-3}$ were obtained. This means that plasma densities in the region crossed by coronal jets, $n_{e\_cor}+n_{e\_jet}$, are on average two times higher than the plasma in the surrounding corona $n_{e\_cor}$.

\subsection{Energy budget}

\indent\indent Given the extension velocity, electron density, and temperature of plasma involved in jets, and combining these parameters with other geometrical and kinematical parameters derived directly by the XRT sequences (like the jet width, length, etc.), it is possible to estimate different possible energies associated with these events. As mentioned in the Introduction, the occurrence of jets is likely due to magnetic reconnections. The energy released during a jet-like eruption may vary from event to event. Following the recent analysis performed by \citet{pucci2013}, the total energy flux $F$ provided to the corona can be expressed as a sum of the fluxes of kinetic energy $F_{kin}$, potential energy $F_{pot}$, enthalpy energy $F_{enth}$, wave energy $F_{wave}$, and radiative energy $F_{rad}$:
\begin{equation}
F = F_{kin}+ F_{pot} + F_{enth} + F_{wave} + F_{rad} \quad [erg\cdot cm^{-2}\cdot s^{-1}].
\label{eq:no4_8}
\end{equation}
\begin{wrapfigure}{r}{0.55\textwidth}
\centering
\includegraphics[width=0.85\linewidth]{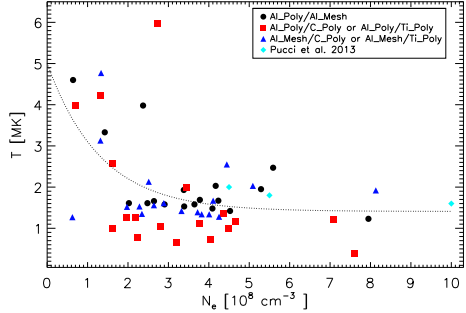}
\caption{Plot showing possible dependence between jet density and temperature. The first three series of data (circle, square, and triangle symbols) are those obtained here with different filter ratios, while other data (diamonds) were obtained by \citet{pucci2013}.}
\label{fig:tempdens}
\end{wrapfigure}
In the above expression, the energy transfer due to thermal conduction is neglected with the following rationale. The timescale for conduction time is expected to be approximately two minutes, a time that is shorter than the average jet lifetime of around ten minutes and much shorter than the jet radiative cooling time 12 to 40 hours \citep[See][]{shimojo2000}. As a result, heat loss by thermal conduction in general should be considered. Nevertheless, as mentioned before, we concluded from our temperature analysis (in agreement also with recent results by \citealt{nistico2011} and \citealt{pucci2013}) that the jets maintain a constant temperature along their height, so the temperature gradient is negligible, and thermal conduction along the jet can be neglected.

The different energy fluxes have been computed as follows. The estimate of the kinetic flux $F_{kin} $ is straightforward:
\begin{equation}
F_{kin} =\frac{1}{2}\rho \mathtt{v}^3\quad [erg\cdot cm^{-2}\cdot s^{-1}]
\label{eq:no4_81}
\end{equation} 
where $\rho = n_e m_p$ represents the mass density of the ejected plasma, and $v$ is the outflow speed of the plasma. The potential flux $F_{pot}$ can be estimated as
\begin{equation}
F_{pot} =\rho \cdot g_\odot \cdot l \cdot \mathtt{v}\quad [erg\cdot cm^{-2}\cdot s^{-1}]
\label{eq:no4_82}
\end{equation}
where $g_\odot$ is the solar gravitational acceleration ($g_\odot=274.13$ $m\cdot s^{-2}$ at the surface), and $l$ is the height (maximum vertical extension of the jet). The enthalpy flux $F_{enth}$ depicts the thermodynamic energy of the jet and is given by\begin{equation}
F_{enth} =\frac{\gamma }{\gamma-1} \cdot p \cdot \mathtt{v}\quad [erg\cdot cm^{-2}\cdot s^{-1}]
\label{eq:no4_83}
\end{equation}
where $\gamma$ represents the ratio of the specific heats (since the coronal plasma can be approximated as a monoatomic gas, we assumed $\gamma=\frac{5}{3}$), and $p$ is the thermal gas pressure ($p=n \cdot k_B \cdot T$ where $k_B$ is the Boltzmann constant). The (Alfv\'en) wave energy flux $F_{wave}$ represents the energy propagating outward with the jet; this is due to Alfven 
waves possibly being excited when the jet-producing reconnection occurs when field lines change their configuration as they relax. Evidence for such Alfven waves appears in the jet movements observed by \citet{cirtain2007}. This quantity can be estimated as
\begin{equation}
F_{wave} =\sqrt{\frac{\rho}{4\pi}}\cdot \xi^2 \cdot B \quad [erg\cdot cm^{-2}\cdot s^{-1}]
\label{eq:no4_84}
\end{equation}
where $B$ is the magnetic field strength. The quantity  $\xi$ represents the amplitude of non-thermal motions in the plasma: here we assumed $\xi$ based on the results by \citet{kim2007}, who found non-thermal velocities in jets ranging from 57 km s$^{-1}$ up to 106 km s$^{-1}$ . The radiative energy flux, $F_{rad}$, represents the energy output due to radiative emission and can be estimated as
\begin{equation}
F_{rad} =n^2_e\cdot \chi T^{\alpha}\cdot l \quad[erg\cdot cm^{-2}\cdot s^{-1}]
\label{eq:no4_85}
\end{equation}
where following analytical approximations by \citet{pucci2013}, we assumed $\alpha=0$ and $\chi=10^{-34.94}$ with $\chi T^{\alpha}$ in W m$^3$. 

The resulting averaged energy fluxes for all the 18 jets analyzed here are $F_{kin}=1.36  \cdot 10^6$ erg cm$^{-2}$s$^{-1}$, $F_{pot}=0.78 \cdot 10^6$ erg cm$^{-2}$s$^{-1}$, and $F_{enth}=7.15 \cdot 10^6$ erg cm$^{-2}$s$^{-1}$, while the resulting $F_{rad}$ component has values that are two orders of magnitude lower than the first three components, hence are neglected.
The average value of the wave energy flux is in the range $F_{wave} = 0.61 \cdot 10^6 - 1.88 \cdot 10^6$ erg cm$^{-2}$s$^{-1}$, depending on the assumed values for non-thermal motion speed, and was computed by assuming a field strength at the base of the jet B = 2.8 G (from \citealt{pucci2013}). The field in polar regions may be even higher, such as 5 G \citep{pucci2013} or even 10 G \citep{ito2010}. The wave energy flux is therefore probably not negligible, but its value is very uncertain due to the uncertainties in the magnetic field and non-thermal velocity values. As a result, in the following discussion we focus on the main sources of energies provided by the jets that we were able to measure independently with XRT data: the kinetic and the enthalpy energy fluxes. The possible energy contribution in coronal holes due to waves will be discussed in the Conclusions.

%

\section{Results}

\indent\indent Here we first investigate possible correlations between different physical parameters of the analyzed jets. Figure \ref{fig:tempdens} shows measured densities plotted against the corresponding measured temperatures for different events. This figure suggests that the temperature of the ejected jet plasma depends on the density of the jet for lower density (hence fainter) events. To increase the statistics, this plot also shows results from \citet{pucci2013}. 

The distribution of points suggests a possible general trend: smaller, fainter events (those with a density lower than $\sim 2 \cdot10^{8}$ cm$^{-3}$) have a higher plasma temperature (higher than $\sim2\cdot10^6$ K) and vice versa. Applying a exponential decay fit of these points gives a Spearman rank coefficient $r_s = 0.21$. This possible dependence is very interesting from the perspective of the estimation of total heating being provided to the corona by jet events: If we consider that fainter (less dense) events are possibly associated with the ejection of hotter plasma, this could lead us to conclude that very high temperature jets (T > 5 MK) are faint enough to have negligible emission with respect to the background corona, thus becoming barely recorded or not at all by XRT. It is therefore possible that the total contribution by polar jets to the coronal heating has been underestimated so far.
\begin{figure*}[t]
\centering
\captionsetup[subfigure]{labelformat=empty}

\subfloat[a]{
\label{fig:fluxratio}
\includegraphics[width=0.45\linewidth]{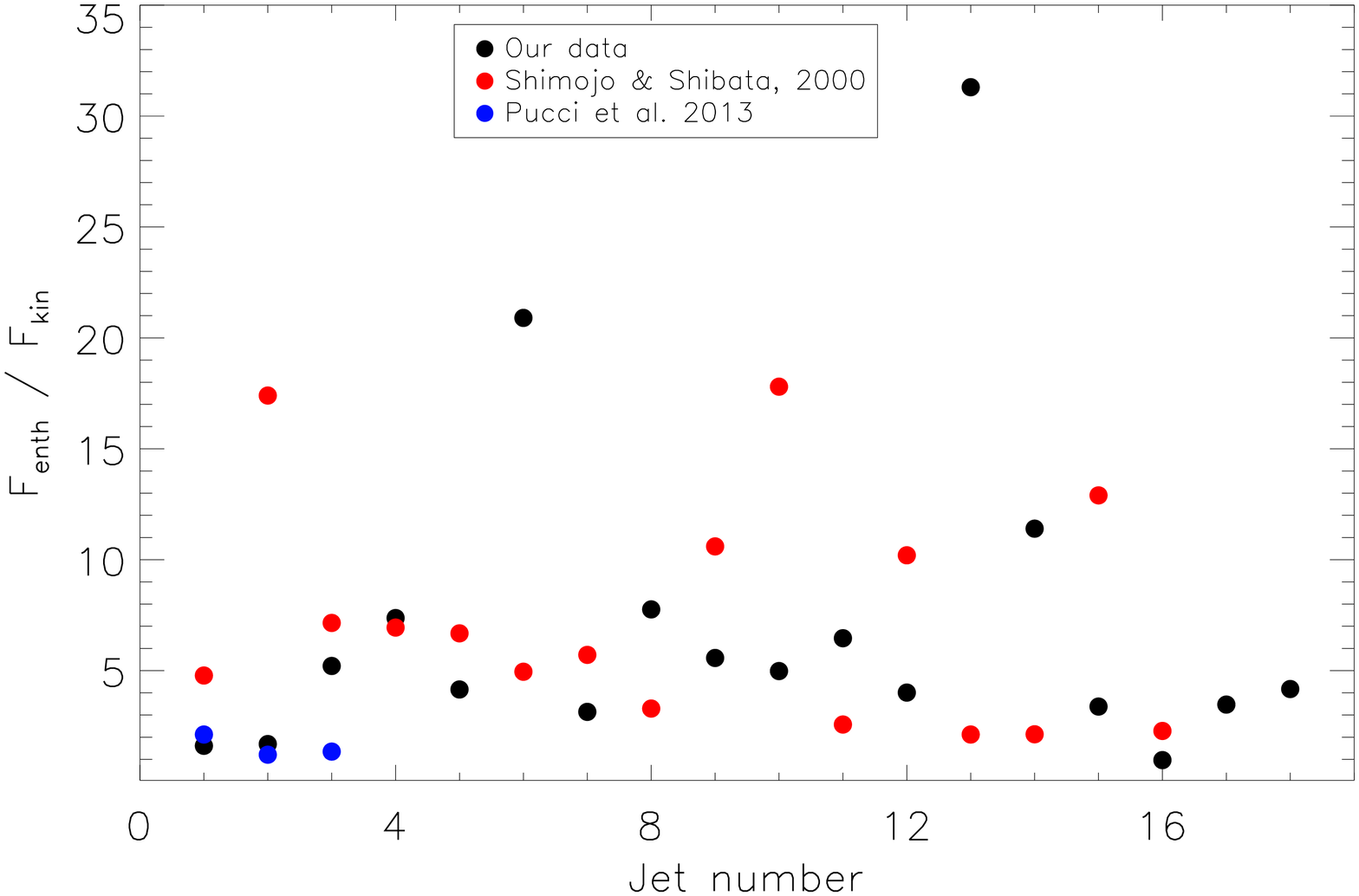}}
\hspace{0.05\linewidth}
\subfloat[b]{
\label{fig:fluxcomp}
\includegraphics[width=0.45\linewidth]{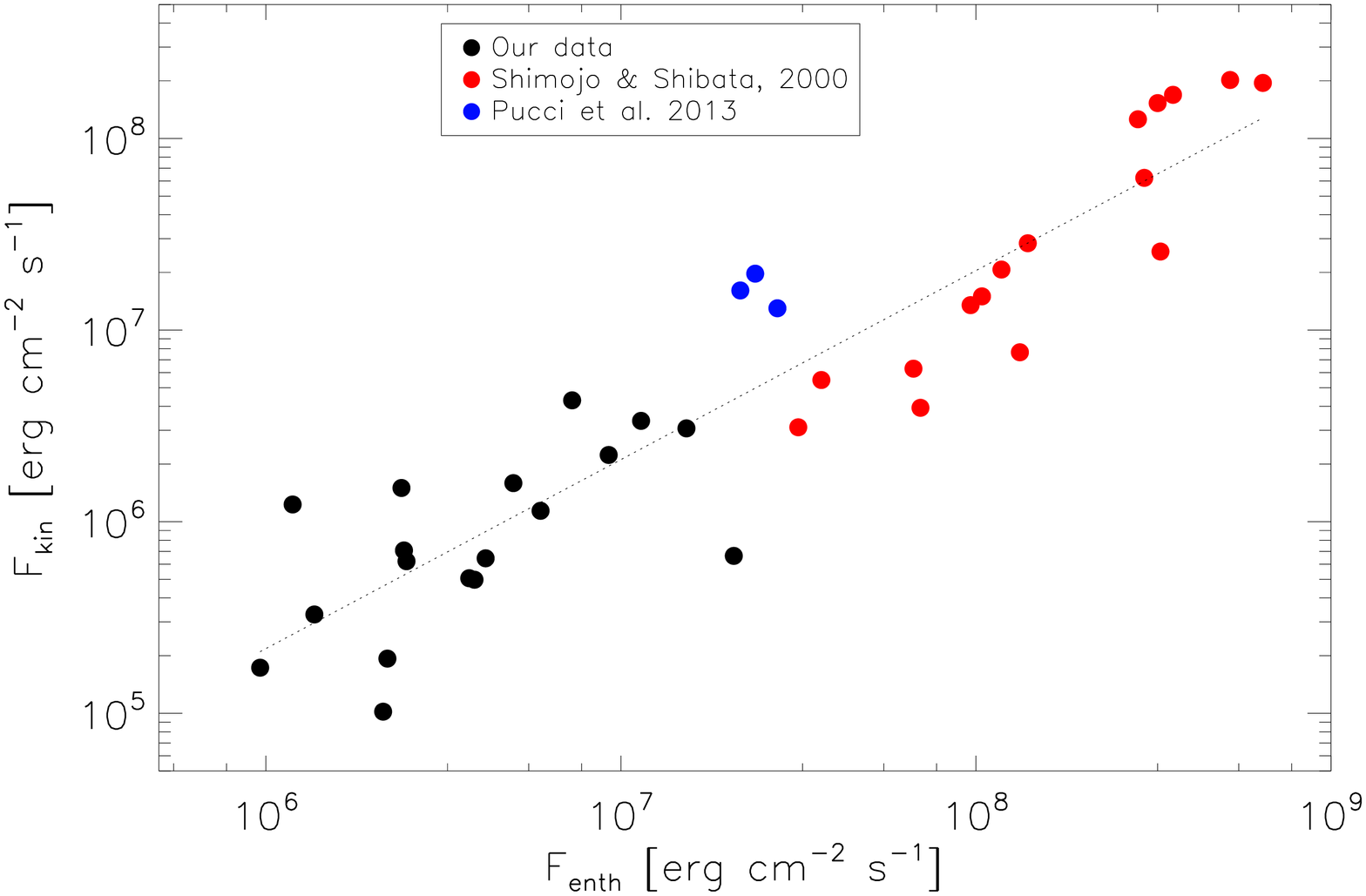}}
\caption{Left panel (a): The ratio between the $F_{enth}/F_{kin}$ fluxes ($y-$axis), computed for the 18 jets studied here ($x-$axis). Right panel (b): Plot of the enthalpy flux (on the $x-$axis, logarithmic scale) versus the kinetic flux ($y-$axis, logarithmic scale). Data from our analysis are plotted here (black circles), together with those (red and blue circles) from the studies performed by \citet{shimojo2000} and \citet{pucci2013}.}
\label{fig:flux}
\end{figure*}

From the estimated energy fluxes, we conclude that the largest portion of energy is provided by the enthalpy flux. The total amount of enthalpy-plus-kinetic energy fluxes has an average value of $F_{tot}=8.5\pm 1.8\cdot 10^6 \text{ }erg\cdot cm^{-2}\cdot s^{-1}$ for all the events analyzed here. In particular, the plot in Figure \ref{fig:fluxratio} shows that the enthalpy energy fluxes for all the events reported in our work are on average more than a factor $\sim3$ greater than the kinetic energy fluxes. As can be seen, these results are also confirmed by data already published in the literature for other events analyzed with other instruments. This implies that the total magnetic energy is not equally converted into kinetic, potential, and thermal energies, as expected from the energy equipartition principle, usually assumed to hold for low $\beta $ plasmas. 
\begin{wrapfigure}{r}{0.55\textwidth}
\centering
\includegraphics[width=0.85\linewidth]{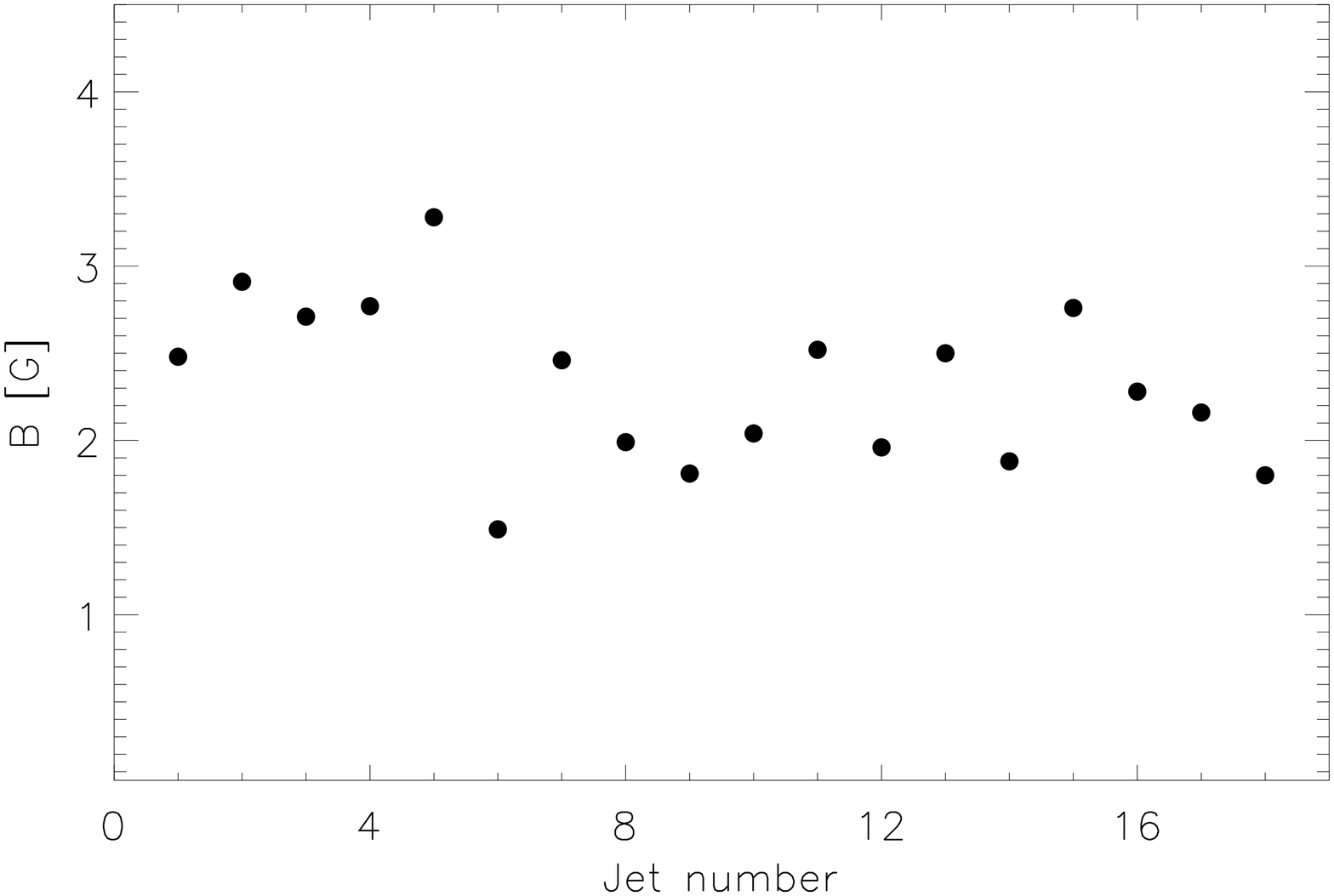}
\caption{Magnetic field strengths calculated via energy conservation ($B_{rec}$) for different jets.}
\label{fig:mf}
\end{wrapfigure}
Moreover, enthalpy and kinetic energy fluxes show a general relative correlation for all the events, as demonstrated by the plot in Figure \ref{fig:fluxcomp}, showing the enthalpy ($x-$axis, logarithmic scale) versus the kinetic ($y-$axis, logarithmic scale) energy fluxes derived for our event and those derived by \citet{shimojo2000} to increase the statistics. This plot shows a clear linear trend (in logarithmic scale): a fit to these points gives a general trend $F_{kin} = (0.26\pm0.02)\cdot {F_{enth} }^{0.98}$ with a Spearman rank coefficient $r_s = 0.92$.

The jet temperatures obtained with the ratios between different filters are on average around $\sim2$ MK, in quite good agreement with previous results in the literature: \citet{pucci2013} used similar XRT data to find jet temperatures of $\sim 1.8$ MK. \citet{culhane2007} studied temperatures and outflow speed of two jets using data from the Hinode EIS Spectrometer and derived temperatures between $0.4$  MK and $ 5$ MK. \citet{nistico2011} studied jets observed in the EUVI and COR1 coronagraph telescope onboard the STEREO mission, and by using the filter ratio technique between the between the EUVI images acquired with the $171\mathring{A}$ and $195\mathring{A}$ bandpass filters, they found values between $0.8$ MK and $1.3$ MK. In another work, \citet{doschek2010} observed jets with Hinode/EIS and found temperatures around 1.4 MK. \citet{madjarska2011} performed a multi-instrument analysis of a jet using SOHO/SUMER, Hinode/XRT and EIS, as well as Stereo EUVI data and derived temperatures of about 12 MK for the bright point at the base of the jet and temperatures in the range of $0.5$ MK and $2$ MK for the jet plasma. All of these measurements for the jet plasma (i.e., not including \citet{madjarska2011} base-bright-point temperatures) are in general agreement, where slight differences most likely result from slight variations in in specific properties of different jets.

 Given the values for the plasma temperature, density, and outflow velocity of the jet, the magnetic field strength at the reconnection point can be estimated by assuming that the whole enthalpy and kinetic energy components are coming from conversion of magnetic energy via the reconnection process, so that
\begin{equation}
\frac{B^2}{8\pi}=\frac{1}{2}\cdot \rho \mathtt{v^2} +n k_B T\Rightarrow B_{rec}=\sqrt{4\pi(\rho \mathtt{v^2}  +2n k_BT)}
\label{eq:no5_1}
.\end{equation}

Applying this formula to data from our 18 jets, we find an average value of $B_{rec} \simeq 2.3\pm 0.5$ G (see Figure \ref{fig:mf}). This value has to be considered as a lower limit estimate because our computation did not include the wave energy or the possible presence of turbulent kinetic energy. This average field strength corresponds, for instance, to the value one may expect at the altitude $h \simeq 5.3 \times 10^3$ km (i.e., 7.3 arcsecs) for the field $B(h) = B_0 (d/h)^2$ \citep[see, e.g.,][]{aschwanden2013} of a small scale dipole with its strength at the photospheric surface $B_0 = 70$ G \citep[as observed at the base of coronal bright points, see e.g.][]{huang2012} buried at the depth $d = 10^3$ km below the solar surface.

\section{Summary and discussion}

A total of 18 polar jets observed by Hinode XRT were selected and analyzed
to derive morphological, kinematical, and physical parameters. The main results of this work can be summarized as follows:
\begin{itemize}
\item The obtained jet outflow speed, averaged over all the events, is $\mathtt{v}_{out}\simeq160\pm 30$ km s$^{-1}$.
\item The average jet temperature was found to be $1.8\pm0.2$ MK.\item Average electron density values of $n_{e\_cor}=1.6\pm 0.2 \cdot 10^8$ cm$^{-3} $ and $n_{e\_cor} + n_{e\_jet}=3.1\pm 0.3 \cdot 10^8$ cm$^{-3}$ were obtained, thus plasma densities in the region crossed by coronal jets were found to be on average two times higher than the plasma in the surrounding corona.
\item The derived densities and temperatures suggest an anti-correlation between these two parameters, with fainter (less dense) jets being hotter and vice versa, thus suggesting that the total energy flux provided to the corona could be underestimated.
\item The largest portion of energy flux provided by jets is due to the enthalpy flux, while the kinetic energy flux is on average about three times smaller. The Alfv\'en wave energy flux may also be substantial, but its value is uncertain. The radiative and conductive losses are negligible.
\end{itemize}

Given the average energy flux for the observed jets, it is also very interesting to estimate the resulting total energy rate provided to the corona due to the jets and to compare this with the energy rate required for coronal heating. Coronal holes typically have a fractional area of 0.02-0.06 of the total solar surface \citep{svalgaard2013}. As a result, if we assume that $A_{CH}$ is the total area covered by coronal holes on the Sun, $A_{jet}=\pi \cdot R^2_{jet}$ is the typical cross sectional area of a jet with observed width $d = 2R_{jet}$, $\tau_{jet}$ is the average jet lifetime, and that $f_{jet}$ is the average polar jet frequency, then the total energy $E_{jet}$ provided by polar jets to the corona can be estimated as
\begin{equation}
E_{jet}=(F_{enth}+F_{kin}+F_{pot}+F_{wave})\frac{A_{jet}}{A_{CH}}\tau_{jet}\cdot f_{jet}
\label{eq:no5_3}
,\end{equation}

where the other energy fluxes have been neglected owing to their small contributions. In this computation, we assume in particular $f_{jet}\simeq 30$ day$^{-1} = 3.47\cdot 10^{-4}$ s$^{-1}$, corresponding to a single coronal hole \citep{savcheva2007}, then $\tau_{jet}\simeq 10$ min = 600 s, $R_{jet}=8\cdot 10^{8}$ cm \citep{savcheva2007}, $A_{CH}\simeq 0.04\cdot A_{sun}\simeq 2.43\cdot 10^{21}$ cm$^{2}$, and $(F_{enth} + F_{kin}+F_{pot}+F_{wave})=(7.15\cdot 10^6+1.36\cdot 10^6+0.78\cdot 10^6+1.24\cdot 10^6)=1.05\cdot 10^7$ erg cm$^{-2}$ s$^{-1}$. Here we used an average value for $F_{wave}$, based on the range of estimates given in Section 3. The resulting total energy rate provided to the corona averaged over the 18 events studied here turns out to be $E_{jet}=0.18\cdot 10^4$ erg cm$^{-2}$ s$^{-1}$. This value is more than two orders of magnitude less than the energy rate required to heat a polar coronal hole, which is estimated to be on the order of $6\cdot 10^{5}$ erg cm$^{-2}$ s$^{-1}$ \citep{withbroe1977}. It therefore appears that the total energy provided by polar jets is not sufficient to heat the solar corona, unless (as mentioned above) we are missing a lot of high-temperature jets that are not observed by XRT or a very significant quantity of turbulent kinetic energy, which is not included in our computation. \citet{moore2011} argue, however, that if spicules are driven by the same processes that drive coronal jets, then, owing to the abundance of spicules, the sum of the energies of coronal jets and those of spicules might be sufficient for powering the corona. Our results also agree with recent work by \citet{yu2014}, who estimated the contribution of coronal jets to the solar wind energy flux and derived an energetic output of $\sim 1.6\%$ compared to the total energy flux of the solar wind.

\section{acknowledgements}

\indent A.R.P. would like to thank the INAF-Turin Astrophysical Observatory, the Faculty of Physics of the University of Bucharest, and the Department of Physics of the University of Turin for funding and support of this work. \\
\indent A.C.S. was supported by funding from NASA's Office of Space Science through the Living With a Star Targeted Research $\&$ Technology Program. A.C.S. also benefited from discussions held at the International Space Science Institute's (ISSI, Bern, Switzerland) International Team on Solar Coronal Jets.\\
\indent Hinode is a solar physics science mission of the Japan Aerospace Exploration Agency (JAXA) in collaboration with the United States and the United Kingdom.

\newpage
\thispagestyle{empty}
\bibliographystyle{plainnat}
\bibliography{bibliography}

\newpage

\begin{table*}
\centering
\caption{\textbf{Appendix: Jet parameters derived with the Al\_Poly/Al\_Mesh ratio.}} \label{Table1}
\begin{tabular}{cccccc}

  \hline
    \\        
     Number  & Dataset & jet width~[km] & Outflow speed~[$km~s^{-1}$] & $T_{e}~[MK]$ & $N_{e}~[10^{8}~cm^{-3}$] \\
             &         &           &                             &                  &                 \\
  \hline
  \\
        1 &1 &  8900$\pm$750   & 180$\pm$16   &  1.9$\pm$0.7  &  3.4$\pm$0.2    \\
        2 &1 &  5200$\pm$750   & 240$\pm$33   &  2.0$\pm$0.4  &  4.2$\pm$0.3    \\
        3 &1 &  8900$\pm$750   & 150$\pm$15   &  1.7$\pm$0.2  &  4.2$\pm$0.1    \\
        4 &1 &  2200$\pm$750   & 120$\pm$1    &  2.5$\pm$0.6  &  5.6$\pm$1.1    \\
        5 &1 &  3700$\pm$750   & 150$\pm$8    &  1.2$\pm$0.4  &  7.9$\pm$2.4    \\
        6 &2 &  5200$\pm$750   & 130$\pm$5    &  4.6$\pm$0.5  &  0.6$\pm$0.1    \\
        7 &2 &  3700$\pm$750   & 160$\pm$14   &  1.4$\pm$0.3  &  4.5$\pm$0.1    \\
        8 &2 &  4400$\pm$750   & 170$\pm$10   &  3.3$\pm$0.4  &  1.4$\pm$0.2    \\                   
        9 &2 &  2200$\pm$750   & 120$\pm$12   &  4.0$\pm$0.6  &  2.4$\pm$0.8    \\
       10 &2 & 12600$\pm$750   & 270$\pm$23   &  1.6$\pm$0.1  &  2.0$\pm$0.4    \\
       11 &2 &  4400$\pm$750   & 130$\pm$29   &  1.7$\pm$0.5  &  3.8$\pm$1.1    \\
       12 &2 &  3700$\pm$750   & 140$\pm$10   &  1.6$\pm$0.6  &  2.5$\pm$0.5    \\
       13 &2 &  6000$\pm$750   & 130$\pm$17   &  1.5$\pm$0.1  &  4.1$\pm$0.8    \\
       14 &2 &  8100$\pm$750   & 100$\pm$13   &  1.7$\pm$0.2  &  2.6$\pm$0.3    \\
       15 &2 &  8900$\pm$750   & 220$\pm$4    &  1.6$\pm$0.3  &  3.6$\pm$0.2    \\
       16 &2 &  8900$\pm$750   & 190$\pm$10   &  1.9$\pm$0.4  &  5.3$\pm$1.6    \\
       17 &2 &  3700$\pm$750   & 140$\pm$5    &  1.5$\pm$0.4  &  3.4$\pm$0.6    \\                   
       18 &2 &  6700$\pm$750   & 120$\pm$4    &  1.6$\pm$0.1  &  2.9$\pm$0.9    \\    
  \hline
\end{tabular}

\end{table*}

\end{document}